# Tuning carrier density and phase transitions in oxide semiconductors using focused ion beams

Hongyan Mei[1*], Alexander Koch[2*], Chenghao Wan[1,3*], Jura Rensberg[2], Zhen Zhang[4], Jad Salman[1], Martin Hafermann[2], Maximilian Schaal[2], Yuzhe Xiao[1], Raymond Wambold[1], Shriram Ramanathan[4], Carsten Ronning[2†], Mikhail A. Kats[1,3,5†]

[1] Department of Electrical and Computer Engineering, University of Wisconsin-Madison, Madison, Wisconsin 53706, USA
[2] Institute of Solid State Physics, Friedrich Schiller University Jena, Jena, 07743, Germany
[3] Department of Materials Science and Engineering, University of Wisconsin-Madison, Madison, Wisconsin 53706, USA
[4] School of Materials Engineering, Purdue University, West Lafayette, IN 47907, USA
[5] Department of Physics, University of Wisconsin-Madison, Madison, Wisconsin 53706, USA

*These authors contributed equally to this work.
[†] E-mail: mkats@wisc.edu , Carsten.Ronning@uni-jena.de

**Abstract:** We demonstrate spatial modification of the optical properties of thin-film metal oxides, zinc oxide and vanadium dioxide as representatives, using a commercial focused ion beam (FIB) system. Using a Ga$^+$ FIB and thermal annealing, we demonstrated variable doping of a band semiconductor, zinc oxide (ZnO), achieving carrier concentrations from $10^{18}$ cm$^{-3}$ to $10^{20}$ cm$^{-3}$. Using the same FIB without subsequent thermal annealing, we defect-engineered a correlated semiconductor, vanadium dioxide (VO$_2$), locally modifying its insulator-to-metal transition (IMT) temperature by range of ~25 °C. Such area-selective modification of metal oxides by direct writing using a FIB provides a simple, mask-less route to the fabrication of optical structures, especially when multiple or continuous levels of doping or defect density are required.

## Introduction

Focused ion beam (FIB) is a well-established technique for high-resolution area-selective milling, deposition, and imaging [1]–[5]. For example, FIB-assisted deposition and milling has been broadly used for applications such as TEM specimen preparation [6], fabrication of electronic and photonic nanostructures [5], [7]–[9], failure analysis [10], and mask repair [11]. Ion implantation using a FIB has also been explored for fabrication of nanoscale devices such as quantum wires [12] and single electron transistors [13] in GaAs/AlGaAs, and Si p$^+$-n junctions for CMOS [14] and CCD [15] applications. Compared to photolithography and e-beam lithography, FIB is a resist-free technique that enables direct etching or deposition of materials with lateral resolution comparable to e-beam lithography (i.e., on the scale of 10 – 100 nm) [5], [12], [13]. In this study, we advance the use of a commercial FIB system to locally modulate the optical properties of metal-oxide via doping or defect engineering. Previously, spatial control of doping [16]–[21] or defect density [22]–[28] has typically been accomplished by implanting ions from ion accelerators through lithographically defined masks, though the FIB has been used to locally tailor optical properties of $Ge_2Sb_2Te_5$ (GST), a chalcogenide-based phase-change material [29], [30]. Here, we extend the use of the FIB to (a) modify the carrier concentration of zinc oxide (ZnO), a wide-bandgap semiconducting oxide, by area-selective doping, and (b) defect-engineer vanadium dioxide ($VO_2$), a prototypical insulator-to-metal transition (IMT) material. The ability to tune the carrier density and phase change behavior via focused ion beam irradiation can enable local patterning of function in nanostructures.

**Tunable carrier concentration in FIB-doped ZnO**

The carrier concentration in most semiconductors can be tuned by orders of magnitude via *in-situ* or *ex-situ* doping processes, resulting in plasma wavelengths from near infrared to far infrared [16], [31]–[34]. Doping can be performed *in-situ* (i.e., during material growth) by tailoring the conditions to introduce dopants during growth processes such as sputtering [35], laser ablation [36], evaporation [37], chemical-vapor deposition [38], etc. In contrast, in *ex-situ* doping techniques, dopants are introduced after material growth, for example via diffusion doping [39], [40], or ion implantation [41]. One advantage of ion implantation is that dopants can be introduced area-selectively, such as by implantation through lithographically defined masks, enabling designer structures, e.g., with plasmonic resonances. For example, we recently used this technique to locally tune the optical properties of silicon to realize all-silicon monolithic Fresnel zone plates and frequency-selective surfaces in the mid and far infrared [16]. In this section, we replace the conventional process of lithography and ion implantation with a FIB-based doping process, realizing mask-free area-selective doping.

We chose zinc oxide (ZnO) as the host material for FIB irradiation. Intrinsic ZnO is transparent from the visible to the mid infrared, and can also be *n*-type doped using gallium (Ga) [34], which is a common ion source in commercial FIB systems. Ga-doped ZnO has been demonstrated as a promising plasmonic material for infrared nanophotonics such as subwavelength waveguides [42]–[44], light-emitting diodes [45], [46], and optical metasurfaces [47]–[49].

The schematic of our FIB-assisted doping process is shown in Fig. 1(a): the ZnO wafer is bombarded by a 30-keV focused Ga ion beam, resulting in the implantation of Ga atoms into the top ~30 nm of the ZnO lattice, but also resulting in lattice damage. A subsequent high-temperature annealing process is necessary for healing the damaged lattice and activating the dopants. As a result, an *n*-type Ga-doped ZnO layer is formed. The penetration depth profile of Ga ions into ZnO [Fig. 1(b)] was estimated using the Monte Carlo code, Transport of Ions in Matter (TRIM) [50], and verified in our samples using Auger electron spectroscopy (AES, Varian Inc.) and X-ray photoelectron spectroscopy (XPS; K-Alpha, Thermo Fisher Scientific) depth profiling (see Section 1 in *Supporting Information*).

The optical properties of metals and metal-like materials can often be approximated using the Drude model [51]. For Ga-doped ZnO, we anticipate that the Drude model should work well in the near-to-mid infrared, with the exception of wavelengths ~20-25 μm, where there is a strong vibrational resonance [52]. In the Drude model, the complex permittivity ($\tilde{\varepsilon}$) is given by

$$\tilde{\varepsilon}(\omega) = \varepsilon_{real} + i\varepsilon_{imag} = \varepsilon_\infty \left(1 - \frac{\omega_p^2}{\omega^2 + i\frac{\omega}{\tau}}\right) \quad (1)$$

$$\omega_p^2 = \frac{n_e q^2}{m^* \varepsilon_0 \varepsilon_\infty}, \quad \lambda_p = \frac{2\pi c}{\omega_p}, \quad \mu = \frac{\tau q}{m^*} \quad (2)$$

where $\varepsilon_\infty$ is the high-frequency permittivity, $\omega_p$ is the screened plasma frequency, which also corresponds to a plasma wavelength ($\lambda_p$), $n_e$ is the carrier concentration and $\mu$ is the carrier mobility determined by the scattering rate ($\tau$), the effective mass of the free carriers ($m^*$), and the unit charge ($q$). The plasma wavelength is the wavelength at which the real part of the permittivity approaches zero, resulting in metal-like behavior at longer wavelengths. As shown in Fig. 1(c), we used the Drude model [Eqs. (1) and (2)] to calculate the complex permittivity ($\tilde{\varepsilon}$) of the Ga-doped ZnO for carrier concentrations from $5 \times 10^{19}$ to $1 \times 10^{21}$ cm$^{-3}$, in which the plasma wavelength is blue-shifted toward the near infrared as the carrier concentration increases.

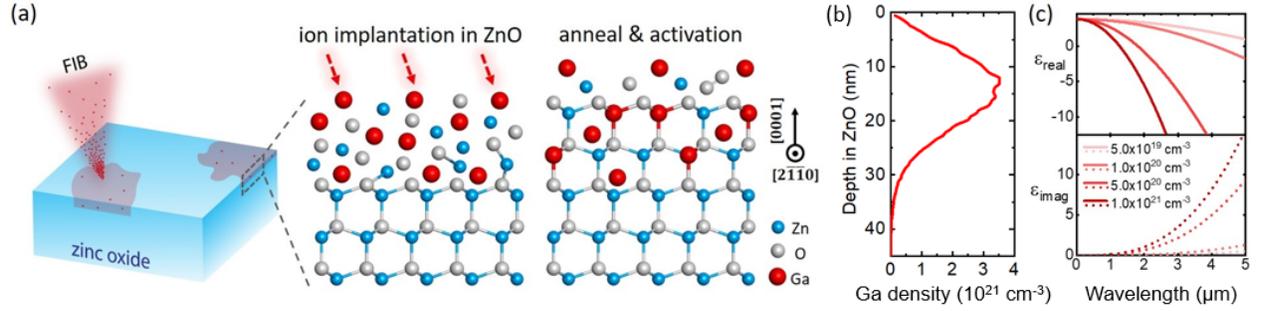

**Figure 1. (a)** Schematic of FIB-assisted doping process: The surface of a single-crystalline ZnO substrate can be doped using a FIB implantation and subsequent high-temperature annealing for activation. **(b)** Depth profile of 30-keV Ga ions impinging into crystalline ZnO, simulated using TRIM. **(c)** Calculated real and imaginary parts of the complex permittivity of Ga-doped ZnO with varying carrier concentrations.

We irradiated several single-crystalline (0001) ZnO substrates (10 × 10 mm$^2$, CrysTec GmbH) with 30-keV Ga ions at room temperature using a commercial FIB system (FEI 600i nanoLab). On each sample, five 200-by-200-µm areas were homogenously implanted with ion fluences of 3.6×10$^{14}$, 6×10$^{14}$, 1.2×10$^{15}$, 3.6×10$^{15}$, and 6×10$^{15}$ cm$^{-2}$ (corresponding to Ga peak concentrations of 0.31, 0.52, 1, 3.1, and 5.2 at.%, respectively), which are close to and above the solid solubility limit of Ga in ZnO [53]–[55]. Note that we irradiated 200-by-200-µm areas to enable far-field optical characterization; in principle, nanometer-scale (10 – 100 nm) lateral resolution can be achieved for the implantation process in a commercial FIB system if diffusion can be avoided. To heal the damaged lattice and activate the Ga dopants, we then performed 40-minute thermal annealing treatments in air of the irradiated samples. Each sample was annealed at a different temperature ranging from 600 to 1000 °C, respectively (complete data and plots for all annealing temperatures can be found in the *Supporting Information* Section 2).

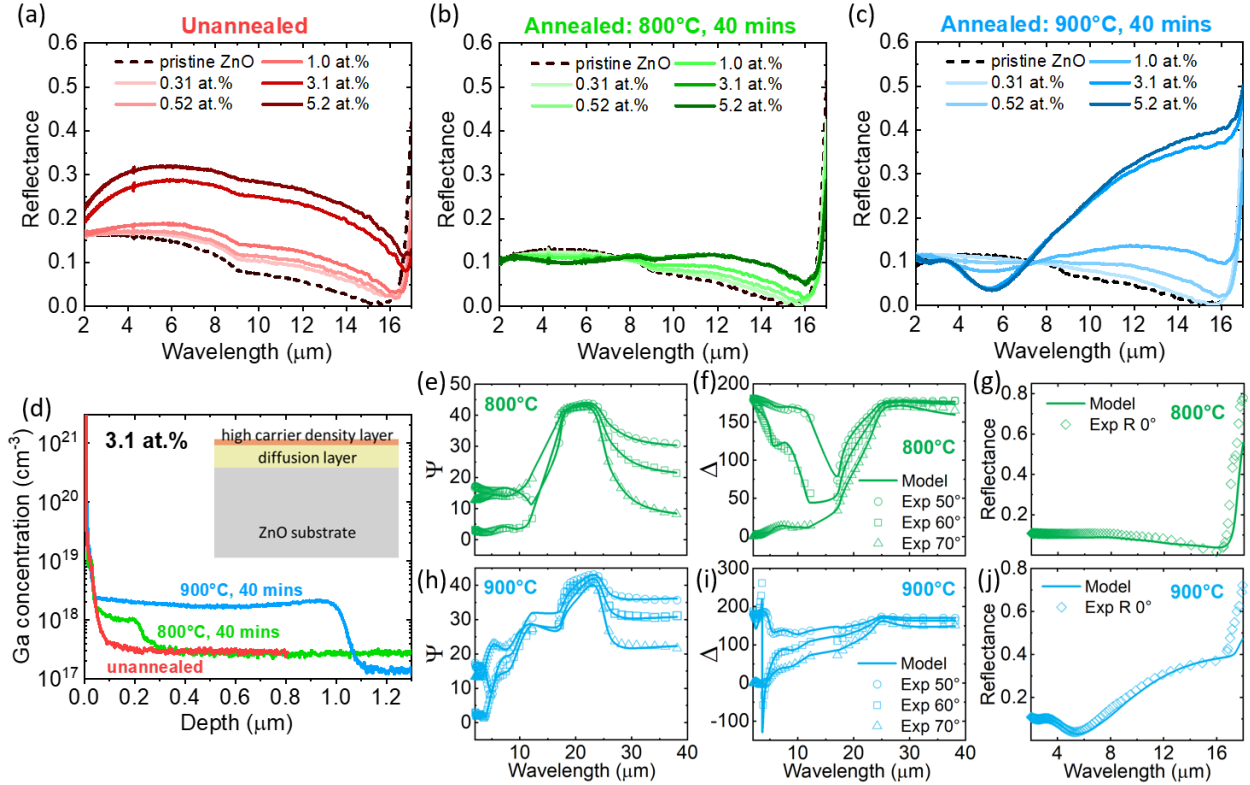

**Figure 2.** Measured normal-incidence reflectance for undoped ZnO, and FIB-irradiated ZnO regions with different ion fluences **(a)** without annealing treatment, **(b)** followed by annealing in air at 800 °C and **(c)** 900 °C, respectively, for 40 mins. **(d)** SIMS depth profiles of the Ga concentration in the samples implanted using an ion accelerator (not a FIB, to enable very large area implantation for infrared ellipsometry and SIMS). The ion energy was identically 30 keV and the peak doping concentration was chosen to be 3.1 at.%. The inset is the schematic showing the three-layer depth profile of thermal annealed FIB-ZnO. **(e-j)** The experimental (discrete points) and model fitted (solid curves) ellipsometric parameters (Ψ and Δ) and normal-incidence reflectance for the 3.1-at.% sample annealed at 800 °C **(e-g)** and for the other one annealed at 900 °C **(h-j)**.

To study the changes in optical properties caused by various doping concentrations and annealing treatments at different temperatures, we performed reflectance measurements on each of these FIB-ZnO regions using a Fourier-transform infrared (FTIR) spectrometer (Bruker Vertex 70) outfitted with an infrared microscope (Hyperion 2000). To study the changes in optical properties caused by various doping concentrations and annealing treatments at different temperatures, we performed reflectance measurements on each of these FIB-ZnO regions using a Fourier-transform infrared (FTIR) spectrometer (Bruker Vertex 70) outfitted with an infrared microscope (Hyperion 2000). For the unannealed Ga:ZnO samples [Fig. 2(a)], we observed increasing reflectance with respect to increasing Ga ion fluence, which we attribute to the partial activation of Ga dopants even without an annealing treatment. Our assumption is supported by a comparison with ZnO substrates that were implanted with Kr ions (see details in Supporting Information Section 5).

As observed in Fig. 2(b–c), the increase of the reflectance versus doping concentration at longer wavelengths (> 8 μm) is as expected due to the activation of dopants. The reduction of the reflectance at shorter wavelengths is likely due to the diffusion of Ga during the annealing treatments which can result in μm-thick doped layers, causing Fabry–Pérot (F-P) fringes at shorter wavelengths (< 8 μm) [Fig. 2(b, c)].

To quantitatively extract physical properties such as carrier concentration and mobility, we performed spectroscopic ellipsometry analysis, which requires centimeter-scale irradiation areas. Therefore, we prepared another set of ZnO substrates irradiated by comparable ion fluences and identical ion energy of 30 keV using an ion implanter, enabling us to homogeneously implant an entire 1-by-1-cm ZnO substrate. Then, we performed spectroscopic ellipsometry (IR-VASE Mark II, J. A. Woollam Co.) measurements for wavelengths from 2 to 20 μm and built a model using ellipsometry analysis software (WVASE, J. A. Woollam Co.) to fit the data. Our assumption about diffusion was confirmed by secondary ion mass spectrometry (SIMS, implemented by Qspec Technology, Inc.) depth profiles as shown in Fig. 2(d). We found ~0.2 μm and ~1 μm plateaus in the 3.1-at.% samples annealed at 800 °C and 900 °C, respectively, which are clear evidence of the diffusion of Ga dopants. Therefore, we built a three-layer model, consisting of a semi-infinite single-crystalline ZnO substrate, one diffusion layer with low carrier concentration, and one top-surface layer with high carrier concentration [inset in Fig. 2(d)]. In our model, we first characterized the pristine ZnO substrate using seven Gaussian oscillators, and for the Ga-doped ZnO (both the top-surface layer and the diffusion layer), an additional Drude oscillator function was added into the oscillator functions of pristine ZnO, to account for the induced carrier concentration due to the doping (see details in *Supporting Information* Section 4). Therefore, the fitting parameters for each sample were the thicknesses, carrier concentrations, and mobilities for the two layers. Note that we kept the seven Gaussian oscillators fixed and only the Drude term was fitted. We used prior knowledge about the thickness of the diffusion layer from SIMS data [Fig. 2(d)] to constrain the fitting for just that parameter; specifically, we constrained the diffusion-layer thickness from 180 nm to 250 nm for the 800 °C annealed sample, and 0.9 μm to 1.1 μm for the 900 °C annealed sample.

**Table 1.** Drude-fitting parameters of the two 3.1-at.% samples annealed at 800 °C and 900 °C, respectively.

| 3.1 at.% Ga:ZnO | High carrier concentration layer | | | Diffusion layer | | |
|---|---|---|---|---|---|---|
| | Thickness (nm) | $n_e$ (cm$^{-3}$) | μ (cm$^2$/V·s) | Thickness (nm) | $n_e$ (cm$^{-3}$) | μ (cm$^2$/V·s) |
| 800°C | 8.0 | 1.25×10$^{20}$ | 19.94 | 205.4 | 1.23×10$^{18}$ | 316.47 |
| 900°C | 8.0 | 2.16×10$^{20}$ | 17.80 | 1008.3 | 2.40×10$^{18}$ | 232.59 |

As shown in Fig. 2(e–j), our model fitted well with the experimental data (Ψ and Δ) acquired using spectroscopic ellipsometry. Note that we excluded the data between 12 and 17 μm in the fitting to avoid non-physical spikes in Ψ and Δ, which result from the low reflectivity of our samples within that wavelength range (more discussion can be found in *Supporting Information* Section 4). The carrier concentration of the 3.1-at.% samples annealed at 800 °C and 900 °C reached 10$^{20}$ cm$^{-3}$ in the top-surface layer, while the carrier concentrations in the diffusion layers underneath are two orders of magnitude lower (Table 1). These fitting results agreed with our SIMS characterizations that the Ga dopants were diffusing from the implantation profile during the annealing process, resulting in a much thicker diffusion layer with a much lower carrier concentration. For most applications, such diffusion layers are unwanted since they trade off patterning resolutions and optical contrast between FIB-irradiated and pristine regions. Since the diffusion layer is highly correlated to the annealing conditions (i.e., annealing temperature and annealing time), plausible methods to decrease the annealing time such as flash lamp annealing [56], [57] and laser annealing [58]–[60] could be useful for suppressing the diffusion.

**Tunable phase-transition characteristics in FIB-engineered VO$_2$**

In the previous section, we demonstrated that optical properties such as carrier density and mobility of an oxide semiconductor (e.g., ZnO) can be locally modified via a simple step of mask-free FIB-assisted ion implantation, followed by a thermal annealing process. We can also use FIB implantation (without annealing) to intentionally introduce structural defects into a material, and the induced defect density can be continuously controlled by varying the ion fluence. Such defect-engineering techniques can be useful for modulating physical properties of materials, especially for strongly correlated electron systems in which electronic properties are very sensitive to changes in the lattice parameters [22], [61]–[73]. In this section, we show that such FIB-assisted defect-engineering can be used to locally modulate the IMT temperature of thin-film VO$_2$, an electron-correlated material that undergoes an IMT at ~70 °C [74], [75] and features an orders-of-magnitude change in carrier density. The IMT temperature of VO$_2$ is determined by the stability of the electron hybridization, which is very sensitive to the strain environment in the thin film [22], [23], [76]–[78]. We previously demonstrated that the IMT temperature can be tuned by introducing structural defects in the VO$_2$ film via high-energy ion irradiation performed using an ion accelerator, where we found the change in optical properties and IMT temperature of VO$_2$ depend on the density of generated defects, but not on the particular ion species (Ar or Cs), and the generated defects introduce more strain to the surrounding and lower the IMT temperature [22].

Here, we show that high-resolution mask-free defect engineering can be accomplished using a commercial FIB system. Similar to the FIB irradiation of ZnO (before annealing), here structural defects are introduced by the collision cascades of impinging ions and lattice atoms (V and O) [Fig. 3(a)], causing changes in the strain environment in the film and thus the IMT temperature is expected to be modulated to different extents depending on the ion fluence.

We deposited a ~50-nm VO$_2$ film on *c*-plane sapphire via magnetron sputtering [75], [79]. Then, twelve 200-by-200-μm regions were irradiated using focused 30-keV Ga ions at room temperature with varying ion fluences up to $2 \times 10^{14}$ cm$^{-2}$, as shown in Fig. 3(b). The density of induced structural defects is proportional to the density of Ga ions implanted into the VO$_2$ film, which we estimated using TRIM simulations [Fig. 3(c)].

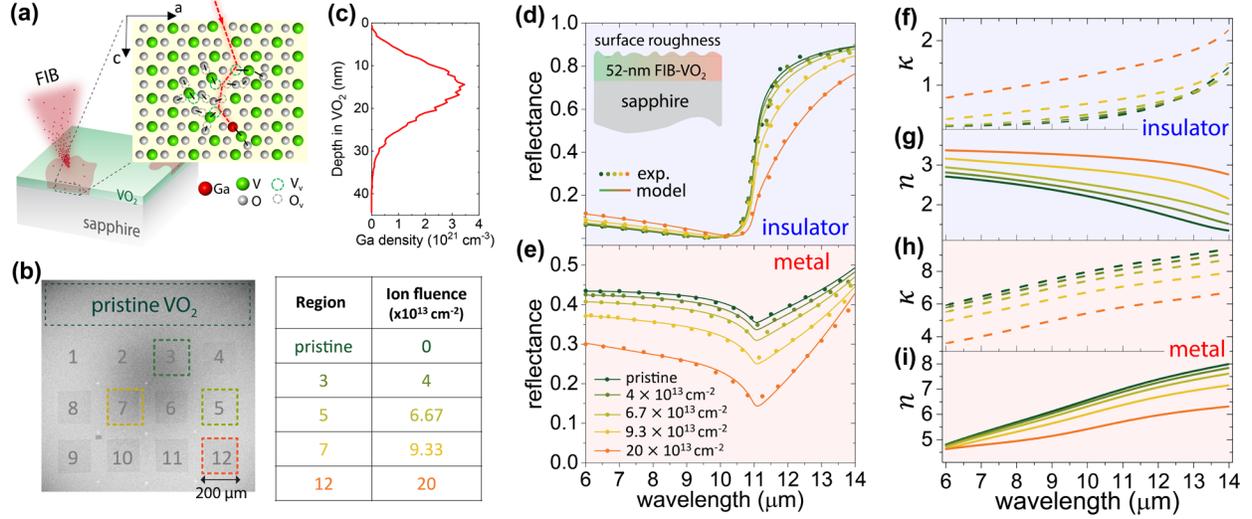

**Figure 3. (a)** Ion irradiation of VO$_2$ using a FIB system, with an inset schematic of the defect-engineering process, showing a collision cascade in the VO$_2$ lattice initiated by an energetic Ga ion. **(b)** SEM image of the FIB-irradiated VO$_2$ regions with the corresponding ion fluences listed. **(c)** Simulated depth profile of 30-keV Ga ions into a VO$_2$ thin film using TRIM. **(d, e)** The symbols are FTIR reflectance measurements on pristine VO$_2$ and regions irradiated with ion fluences of 4 x 10$^{13}$, 6.7 x 10$^{13}$, 9.3 x 10$^{13}$, 20 x 10$^{13}$ cm$^{-2}$, for temperatures at 25 °C (all regions in pure insulating phases) and 100 °C (all regions in pure metallic phases), respectively. The solid curves are the model fits to the FTIR measurements, where the underlying model was created based on ellipsometry of pristine VO$_2$. The insulator-phase **(f, g)** and metal-phase **(h, i)** refractive indices extracted from the fittings shown in (d) and (e).

To investigate the irradiation-induced changes in the optical properties of the pure insulator- and metal-phase VO$_2$, we first performed reflectance measurements on each of these FIB-irradiated VO$_2$ regions using our FTIR spectrometer with microscope, for temperatures of 25 °C (i.e., VO$_2$ in the pure insulating phase for all the irradiated regions) and 100 °C (i.e., VO$_2$ in the pure metallic phase for all the irradiated regions), as shown in Fig. 3(d) and (e). Then, we fitted the measured reflectance by adjusting the parameters of a model that we previously built to characterize refractive indices of intrinsic thin-film VO$_2$ [75]. As shown in the inset of Fig. 3(d), the model consisted of a semi-infinite anisotropic *c*-plane sapphire [75], a VO$_2$ layer, and surface roughness (50% air + 50% of the material underneath). For the insulating phase, the dielectric function of the VO$_2$ layer is a series of Lorentzian oscillators. For the metallic phase, we also used Drude functions to capture the contribution of the free carriers (more details can be found in ref. [75]). The thicknesses of VO$_2$ and surface roughness were set to 52 nm and 5 nm, respectively, based on SEM imaging of the cross section (*Supporting Information* Section 5). We were able to fit our reflectance measurements [Fig. 3(d, e)] by only adjusting the line shapes, amplitudes, and spectral positions of the Lorentz and Drude functions. Therefore, the complex refractive indices of VO$_2$ for different ion fluences can be extracted.

Then, we investigated FIB-induced modulation of the IMT temperature and width by a combination of temperature-dependent FTIR reflectance measurements and effective-medium theory, as schematically shown in Fig. 4(a). FTIR reflectance measurements were performed on all irradiated regions for temperatures increasing from 10 to 120 °C, in steps of 2 °C. We observed that the phase transition shifted to lower temperatures as the ion fluence increased, which agrees with our previous observations for defect-engineered VO$_2$ irradiated using an ion accelerator [22]. To quantitatively study the changes of IMT characteristics with respect to the FIB fluence, we used the Looyenga effective-medium theory formalism [60] to approximate the refractive indices of the irradiated VO$_2$ at intermediate temperatures [75], [80]:

$$\tilde{\varepsilon}_{eff}^{1/3} = (1-f)\tilde{\varepsilon}_i^{1/3} + f\tilde{\varepsilon}_m^{1/3} \qquad (3)$$

where $\tilde{\varepsilon} = \tilde{n}^2 = (n+i\kappa)^2$ is the complex dielectric function of VO$_2$ and $f$ is the temperature-dependent volume fraction of the metal-phase VO$_2$ domains within the film. The co-existence of insulating and metallic domains can be understood as a first-order equilibrium distribution, and therefore $f(T)$ can be expressed as [22], [81]:

$$f(T) = \frac{1}{1+\exp\left[E/k_B\left(1/T - 1/T_{IMT}\right)\right]} \qquad (4)$$

where $E$ is an energy scale that determines the sharpness of the IMT (i.e., inversely proportional to the IMT width). $T_{IMT}$ is the temperature where 50% of VO$_2$ transformed to the metallic phase in a heating process.

For given $E$ and $T_{IMT}$, we used Eq. 3 and 4 to obtain the temperature-dependent refractive indices and then calculated the optical reflectance of each irradiated region using the transfer-matrix method. As shown in Fig. 4(a), by sweeping $E$ and $T_{IMT}$, we achieved good agreement between the calculation (solid curves) and FTIR measurements (dotted lines). The fitted IMT temperature and width as a function of ion fluence are shown in Fig. 4(b). The IMT width is defined to be the temperature interval between where 3% of VO$_2$ is in the metallic phase and where 97% of VO$_2$ is in the metallic phase. Note that due to the hysteresis in VO$_2$, the value of $T_{IMT}$ is different for heating and cooling [75], [82]. Once $f(T)$ was determined, we were able to obtain the temperature-dependent refractive indices across the IMT for each irradiated region, as shown in Fig. 4(c). Here, we only plot the results for a single wavelength ($\lambda$ = 9 µm) to better show the evolution of the refractive indices with respect to both the temperature and the FIB fluence. The full dataset for wavelengths from 6 to 14 µm can be found in *Supporting Information* Section 6.

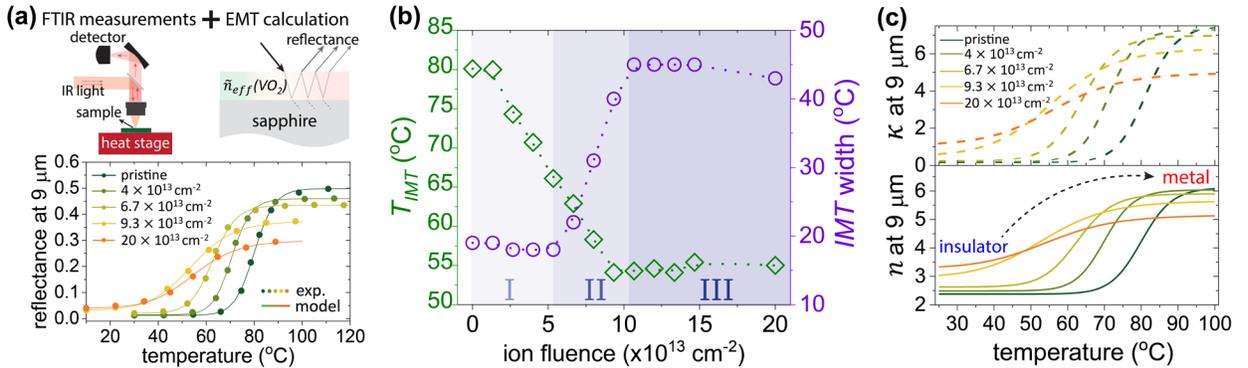

**Figure 4.** (a) Temperature-dependent optical characterization of the FIB-irradiated VO$_2$. First, we measured temperature-dependent reflectance across the IMT for each irradiated region. Then, we applied effective-medium theory to approximate the refractive indices at intermediate temperatures and calculate the temperature-

dependent reflectance. By sweeping the parameters of $T_{IMT}$ and $E$ —which determine the IMT width and temperature, respectively—in Eq. (4), we found the best fit between the FTIR measurements and calculation, enabling us to extract the IMT temperature and width for each irradiation ion fluence, as plotted in **(b)**. **(c)** Extracted temperature-dependent refractive indices of the defect-engineered VO₂ irradiated by different ion fluences. Here we plot the results for a single wavelength of 9 μm to clearly show the evolution of refractive-index values versus temperature and ion fluence.

As shown in Fig. 4(b), there are three distinct ion fluence regimes (labeled in the figure as I, II, and III), in which the IMT characteristics evolve differently. For ion fluences $< 5 \times 10^{13}$ cm$^{-2}$, the IMT temperature gradually decreases with fluence, with a reduction of ~15 °C for $5 \times 10^{13}$ cm$^{-2}$ and no substantial changes in either the refractive index of the two pure phases or in the IMT width. At higher ion fluences between $5 \times 10^{13}$ and $1.1 \times 10^{14}$ cm$^{-2}$, we observed that the IMT temperature could be further shifted to lower temperatures, but the shift was accompanied by a significant broadening in the IMT width and a reduction in the refractive-index contrast between the two pure phases [Fig. 4(c)].

We attribute such distinct phenomena to the different defect morphologies induced by different levels of ion fluence. At low ion fluences, the impinging ions mostly cause point defects that can reduce the transition temperature due to local compressive strain [22], [78]. Such point defects are in much smaller than the probing wavelengths of our FTIR measurements. The strain induced by the point defects can be redistributed and partially relaxed at room temperature after irradiation [83], [84], resulting in a homogeneous strain environment in the film. This understanding is consistent with the lack of broadening in the IMT width in the low-fluence irradiation regions in Fig. 4. As the ion influence increases, point defects are expected to accumulate and form nanometer-sized defect complexes that can affect the IMT temperature in microscopic scales, thus resulting in apparent broadening of the IMT in regions irradiated with high fluence.

When the ion fluence surpasses ~$1.1 \times 10^{14}$ cm$^{-2}$, both the IMT temperature and width became constant versus the increasing ion fluence, likely due to the limited penetration depth of the 30-keV Ga ions in the VO₂ film. At these high ion fluences, we expect the density of the induced structural defects complexes to saturate (i.e., complete amorphization occurs [22]) within the depth of ~30 nm from the VO₂ surface, while leaving a less-affected VO₂ layer underneath, as shown in our TRIM simulation [Fig. 3(c)].

**Summary**

We have shown that the optical properties of two oxide materials, zinc oxide (ZnO) and vanadium dioxide (VO₂), can be locally modulated by doping or defect engineering using a commercial focused ion beam (FIB) with gallium ions. Using the FIB, we modified the carrier concentrations in initially undoped ZnO, reaching carrier concentrations as high as $10^{20}$ cm$^{-3}$, and reduced the temperature of the insulator-to-metal transition (IMT) in VO₂ by as much as ~25 °C. The FIB process does not require any lithography or masking, and only requires one additional annealing step in the case of doping. Due to the versatility of commercial FIBs, this technique can be used to modify and engineer materials with high resolution even for the case of irregularly shaped materials where conventional lithography is challenging. The ability to dope and defect-engineer certain oxides using a commercial FIB provides functionalities beyond the more-common FIB milling and deposition, and may enable the direct fabrication of a broader range of infrared devices based on semiconducting oxides.

## Acknowledgement

M.K. is supported by the Office of Naval Research (ONR, N00014-20-1-2297). C.R. is supported by the Deutsche Forschungsgemeinschaft (DFG) through grant Ro1198/21-1 and by a collaborative exchange program of the Deutscher Akademischer Austauschdienst (DAAD) through grant 57386606. S.R. acknowledges AFOSR grant FA9550-18-1-0250. The authors gratefully acknowledge use of facilities and instrumentation at the UW-Madison Wisconsin Centers for Nanoscale Technology (wcnt.wisc.edu) partially supported by the NSF through the University of Wisconsin Materials Research Science and Engineering Center (DMR-1720415).

## Data Availability

The data that support the findings of this study are available from the corresponding author upon reasonable request.

# Supporting Information:
# Tuning carrier density and phase transitions in oxide semiconductors using focused ion beams

Hongyan Mei[1*], Alexander Koch[2*], Chenghao Wan[1,3*], Jura Rensberg[2], Zhen Zhang[4], Jad Salman[1], Martin Hafermann[2], Maximilian Schaal[2], Yuzhe Xiao[1], Raymond Wambold[1], Shriram Ramanathan[4], Carsten Ronning[2†], Mikhail A. Kats[1,3,5†]

[1] *Department of Electrical and Computer Engineering, University of Wisconsin-Madison, Madison, Wisconsin 53706, USA*
[2] *Institute of Solid State Physics, Friedrich-Schiller-Universität Jena, Jena, Thuringia 07743, Germany*
[3] *Department of Materials Science and Engineering, University of Wisconsin-Madison, Madison, Wisconsin 53706, USA*
[4] *School of Materials Engineering, Purdue University, West Lafayette, IN 47907, USA*
[5] *Department of Physics, University of Wisconsin-Madison, Madison, Wisconsin 53706, USA*
*These authors contributed equally to this work.*
[†]*E-mail: mkats@wisc.edu , Carsten.Ronning@uni-jena.de*

## Section 1. X-Ray Photoelectron Spectroscopy (XPS) and Auger Electron Spectroscopy (AES) of Ga:ZnO

### X-Ray Photoelectron Spectroscopy (XPS)

XPS depth profiling (K-Alpha, Thermo Fisher Scientific) was performed to investigate the chemical binding states and the relative atomic composition in Ga:ZnO. Four samples were investigated: intrinsic zinc oxide (ZnO), as-implanted Ga:ZnO (i.e., samples without annealing treatments) and two Ga:ZnO samples annealed at 900 °C and 1000 °C for 40 minutes. All Ga:ZnO samples were implanted with a ion fluence of $6 \times 10^{15}$ cm$^{-2}$, which corresponds to Ga peak concentration of 5.2 at.% at an ion range of ~14 nm and a straggling of ~6 nm, which was calculated using an open-source Monte Carlo code, Transport of Ions in Matter (TRIM) [S1].

First, we performed a survey scan on the as-implanted Ga:ZnO to estimate surface contaminations and detectable elements. Then, we performed highly resolved XPS measurements on each detectable element. The Shirley background subtraction and the peak fitting with asymmetric Lorentzian or Gaussian line shape profiles were executed on the high-resolution XPS spectra for chemical state analysis (implemented in the CasaXPS [S3]). We report XPS depth profiles of Ga2p, Zn2p, O1s and C1s. The depth profiles were realized with a stepwise etching process using Ar ions with an etching rate of 0.1 nm/s. We first removed the surface contaminations with an 50-second etching and then started with a depth profiling of Ga:ZnO for 600 seconds in total (etching for 10 seconds per cycle, 31 scans in total) to get depth resolved XPS spectra including both the Ga-implanted and non-implanted region.

The survey scan of the as-implanted Ga:ZnO surface is shown in Fig. S1, in which peaks of Zn, O, C and Ga were detected. The spectrum shows nontrivial surface contamination with C and O due to the long exposure in air.

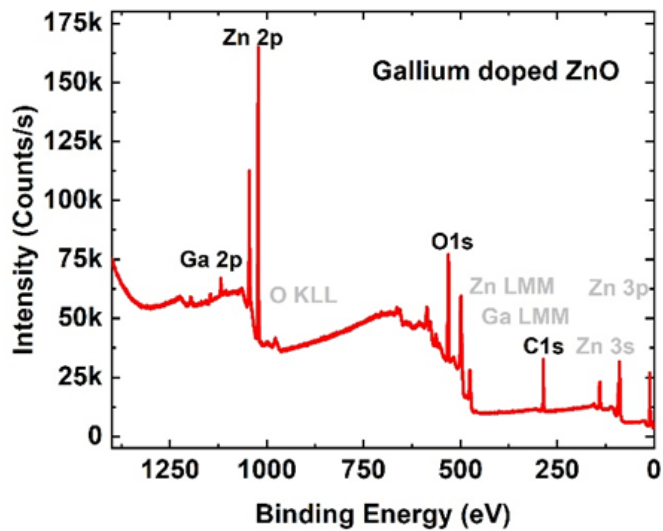

**Figure S1.** XPS survey scan of as-implanted Ga:ZnO sample surface with binding energies for several peaks.

Fig. S2(a) shows the highly resolved C1s spectra with three components, which describes most probably adventitious carbon contamination on the Ga:ZnO surface [S2]–[S4]. The main peak I (285.7 eV) represents pure carbon (C-C), peak II (287 eV) stands for carboxyl groups (C-O) and peak III (289.7 eV) describes carbonates (O-C=O) [S2]–[S4]. All Ga:ZnO samples exhibit a similar carbon contamination at the surface, which need to be removed by $Ar^+$ etching for element quantification of Ga:ZnO. Fig. S2(a) includes the peak intensity of C1s in function of etching depth. In total we etched for 600 s, i.e., 60-nm deep from surface into the substrate. After 50 seconds of etching we could still see a small peak for pure carbon (at the location of the blued dashed line), which was very likely due to the recoil implantation.

After 50 seconds of etching, we started depth profiling for the element quantification and chemical binding state analysis, in which the Zn2p- [Fig. S2(b)], O1s- [Fig. S2(c)] and Ga2p- [Fig. S2(d)] peaks were considered. As shown in Fig. S2(b) the peak composition (i.e., the binding state) of Zn2p does not change with respect to the depth. The binding energies of 1022.5 eV and 1045.5 eV are corresponding to $Zn2p_{3/2}$ and $Zn2p_{1/2}$ within the tetrahedral coordination of crystalline ZnO [S2], [S3], [S5], [S6]. The $Zn2p_{3/2}$ peak shifted from 1022.5 eV to 1022.2 eV during 600 seconds of etching, indicating a small portion of Zn-O bonds were either broken or substituted by Ga-O bonds after the Ga ion implantation.

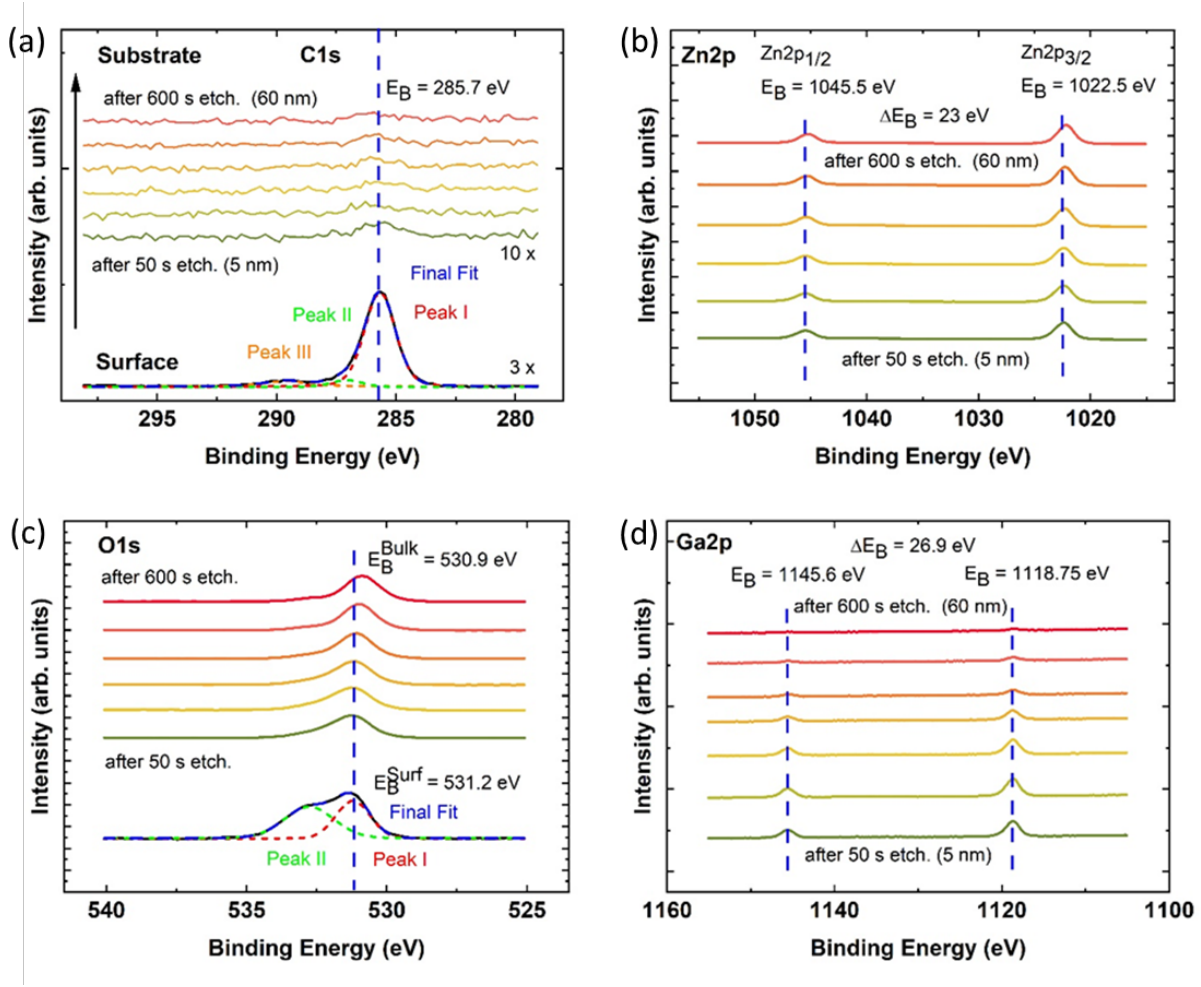

**Figure S2. (a)** Surface and depth profile of C1s spectra of the as-implanted Ga:ZnO sample. The depth profile (green-to-red lines) is a series of highly resolved XPS spectra measured in a sequence of 100-second etchings. For a better view the C1s spectra are scaled up. **(b)** Depth profile of Zn2p spectra of the unannealed Ga:ZnO substrate. **(c)** Surface and depth profile of O1s spectra for the as-implanted Ga:ZnO. **(d)** Depth profile of Ga2p spectra of the as-implanted Ga:ZnO.

As shown by the solid blue curve in Fig. S2(c), the surface XPS spectra of the O1s can be fitted by two Gaussian peaks: peak I at 531.2 eV that is attributed to Zn-O bonding and peak II at 532.7 eV that indicates the presence of O-H bonds induced by the surface contamination [S2], [S3], [S5], [S6]. After the initial 50-second etching, Peak II disappeared because the surface contamination was removed. Then, within the further etching, Peak I shifted by ~0.3 eV, similar as that of the $Zn2p_{3/2}$ peak, which is due to the lattice damage during the Ga ion implantation.

In the depth profile of the Ga2p spectra [Fig. S2(d)], the bonding energies are found to be 1118.75 eV for $Ga2p_{3/2}$ and 1145.6 eV for $Ga2p_{1/2}$, which are 2 eV higher than the binding energies in Ga metal, indicating a binding between Ga and O in the ZnO substrate [S1], [S2], [S4]–[S6]. On the other hand, there is no peak at 1117.5 eV [S2], [S7], which is the typical bonding energy of Ga-O in gallium oxide($Ga_2O_3$), indicating no macroscopic agglomerates of $Ga_2O_3$ formed during the implantation or the concentration of $Ga_2O_3$ is less than the XPS-detection limit. In addition, Ga2p peak intensity is decreasing as a function of depth, which agrees well with our TRIM simulation results shown in Fig. 1(b) in the main text.

Fig. S3 plots the concentration of Ga as a function of depth for the as-implanted Ga:ZnO and the samples annealed at 900 °C and 1000 °C. In general, the Ga concentration in the annealed samples were found to be lower than that of the as-implanted sample, indicating the thermal diffusion of Ga from the surface into the bulk ZnO during the annealing process. Furthermore, such diffusion is highly dependent on the annealing temperature, as we found that the Ga concentration left at the surface region in the sample annealed at 1000 °C (blue curve) is much less than that of the sample annealed at 900 °C (red curve). We also observed several subpeaks of Ga concentration in the annealed samples (two in the red curves and one in the blue curve), indicating a re-accumulation process of Ga that is also depending on annealing conditions (i.e., temperature and time). Though the dynamic of such re-accumulation has not been fully understood, we believe it can be avoided by a more-sufficient annealing treatment (i.e., annealing with either sufficiently high temperature or sufficiently long time).

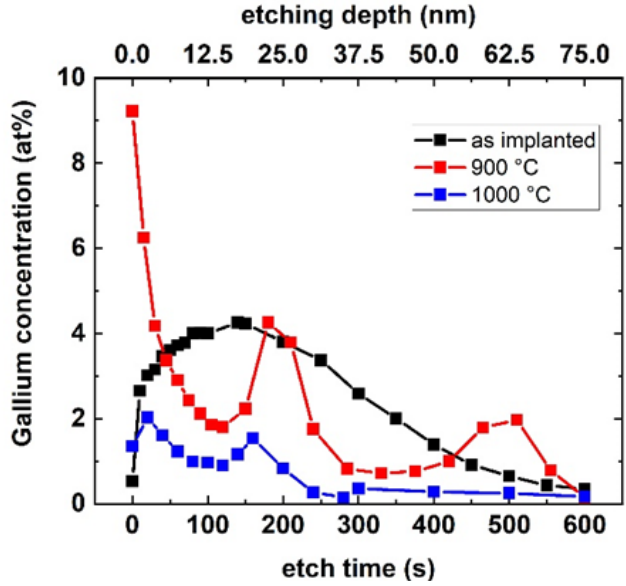

**Figure S3.** Depth profiles of Ga concentration for the as-implanted Ga:ZnO and Ga:ZnO samples annealed at 900 °C and 1000 °C for 40 minutes.

**Auger Electron Spectroscopy (AES)**

We also performed AES measurements (Varian Inc.) on the same as-implanted Ga:ZnO to confirm the depth profile of the Ga concentration, by comparing the results to those of the TRIM simulation and the XPS measurements. The depth profiling was done via a stepwise etching process using krypton (Kr). The total etching process resulted in a removal of 68-nm material. The etching rate was 0.12 nm/s with the step size of 30 seconds.

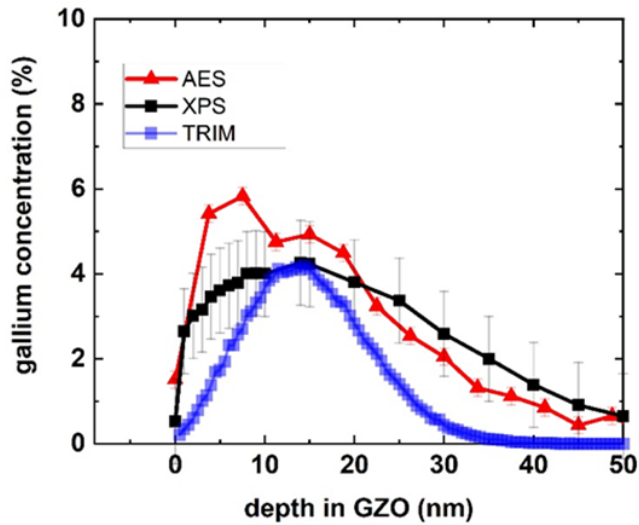

**Figure S4.** Comparison between AES, XPS, and TRIM results of the depth profiling of Ga concentration in the as-implanted Ga:ZnO sample that was implanted by 30-keV Ga ions with a fluence of $6 \times 10^{15}$ cm$^{-2}$.

Comparison among the results of AES, XPS, and TRIM simulation is summarized in Fig. S4. As discussed in the main text, TRIM simulation predicts a Gaussian-like depth profile of the Ga distribution with the peak concentration, $c_{Ga}$, of ~5.2 at.%, which matches well with that of the XPS measurements. The measured depth profiles by XPS or AES featured broader Ga distribution ranges than that of the TRIM simulation which is probably due to slight thermal diffusion of Ga caused by the generated heat during the implantation.

We estimated the total amount of implanted Ga by calculating area under the concentration curves. For the TRIM simulation, the total implanted Ga concentration, $c_{Ga}$, is $6 \times 10^{22}$ cm$^{-3}$. For the experiments, we got $c_{Ga}$ of $1.23 \times 10^{23}$ cm$^{-3}$ from AES results and $1.18 \times 10^{23}$ cm$^{-3}$ from XPS results, which agree well with each other.

## Section 2. FTIR reflectance of Ga:ZnO samples annealed at different temperatures

Several single-crystalline ZnO substrates were irradiated using a 30-keV focused Ga ion beam at room temperature. On each sample, we homogenously implanted five 200-by-200-μm areas with different ion fluences that correspond to Ga concentrations of 0.31, 0.52, 1, 3.1, and 5.2 at.%. Then, the samples were thermally annealed in air for 40 minutes at temperatures of 600, 700, 800, 900 and 1000 °C. Optical reflectance of these FIB-ZnO regions were measured using our FITR microscope and the results are plotted in Fig. S5, in which three major trends were found: a) For the unannealed samples [Fig. S5(a)], the increase of the reflectance with respect to the increasing ion fluence is likely due to the introduction of Ga ions and structural defects at the ZnO surface. After the annealing at lower temperatures [i.e., Fig. S5(b and c)], the reflection magnitudes and line shapes of those FIB-ZnO regions "recovered" to that of the undoped ZnO spectrum, which could be due to multiple effects including healing of the induced defects, the formation of Ga-O bonds, and the diffusion of Ga ions into the bulk ZnO. b) In Fig. S5(d, e, and f), we observed an obvious increase in the reflectance with respect to the increasing Ga concentration at longer wavelengths (> 8 μm), which is attributed to the activation of dopants during the annealing process. c) On the other hand, at shorter wavelengths (< 8 μm) the reduction of reflectance is likely due to the diffusion of Ga during the annealing treatments, resulting in μm-thick less-doped layers that caused Fabry–Pérot fringes.

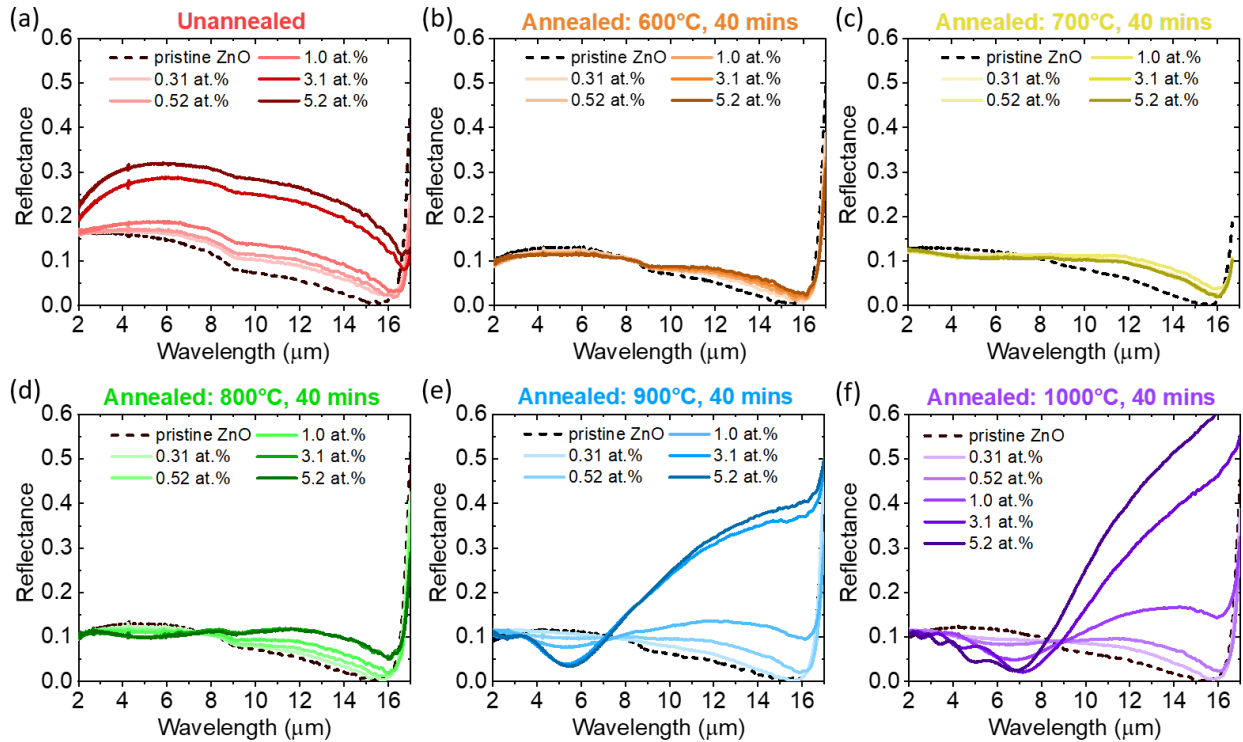

**Figure S5.** Measured near-normal-incidence reflectance for undoped ZnO, and areas irradiated with different ion fluences followed by **(a)** no annealing treatment, **(b-f)** annealing in air for 40 minutes at temperatures of 600, 700, 800, 900, and 1000 °C, respectively.

**Section 3. Comparison between ZnO implanted by Kr and Ga ions**

In order to understand the increasing reflectance with the increasing Ga ion fluences in the unannealed samples, we implanted ZnO with 30-keV Krypton (Kr) ions using an ion implanter, for ion fluence ranging from $1.6 \times 10^{15}$ cm$^{-2}$ to $1.6 \times 10^{16}$ cm$^{-2}$, resulting in similar defect concentration distribution as that of Ga ions. Kr is a noble gas with an atomic mass of 84, which is slightly heavier than Ga with an atomic mass of 69. Therefore, the ion fluences we used for Kr is a little bit smaller than for Ga. Figure S6(a) shows the generated vacancy profiles at 30 keV versus the target depth, according to TRIM calculations [S1].

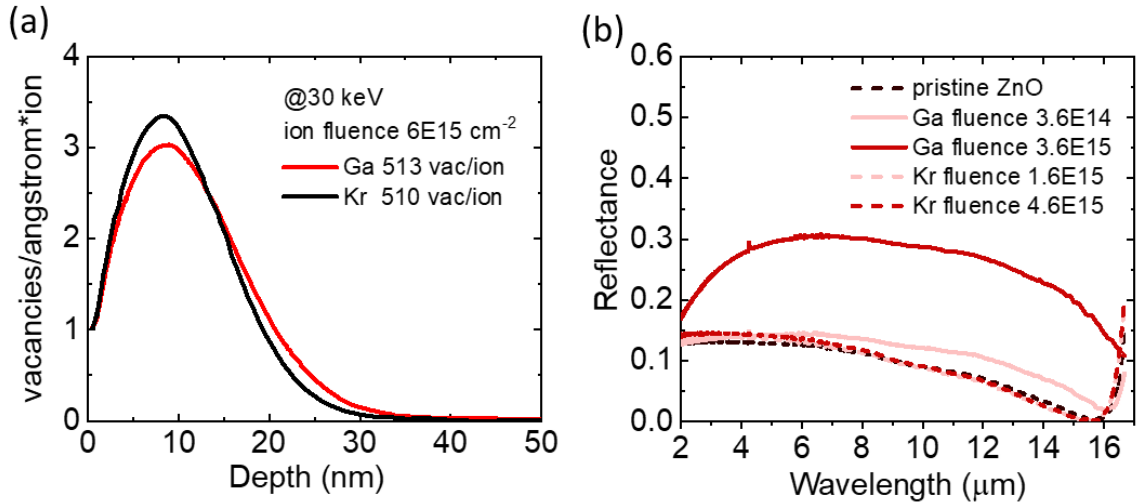

**Figure S6. (a)** TRIM-calculated vacancy distribution of Ga and Kr implantation at 30 keV, both with an ion fluence of $6 \times 10^{15}$ cm$^{-2}$. **(b)** Reflectance of Kr-implanted single-crystalline ZnO and Ga-implanted single-crystalline ZnO, both without annealing.

Figure S6(b) presents the FTIR reflectance of ZnO implanted with Kr and Ga ions using similar implantation conditions. The reflectance of the Kr-implanted ZnO is close to that of the pristine ZnO, indicating that any implantation-induced defects do not affect the optical properties. In contrast, the reflectance of ZnO significantly increased after the Ga implantation, which we attributed to some Ga dopant activation even without annealing, resulting in a thin layer of highly doped Ga:ZnO.

**Section 4. Surface morphology of Ga:ZnO samples annealed at different temperatures**

We took SEM images of the FIB-ZnO areas irradiated with the fluence of $6\times10^{15}$ cm$^{-2}$ and annealed at different temperatures (Fig. 5). For the samples annealed at high temperatures [Fig. S5(c and d)], the process of the recrystallization was observed. The hexagonal features in Fig. 5(d) are likely the wurtzite ZnO [S8], which is the most thermally stable form for single crystalline ZnO in ambient air.

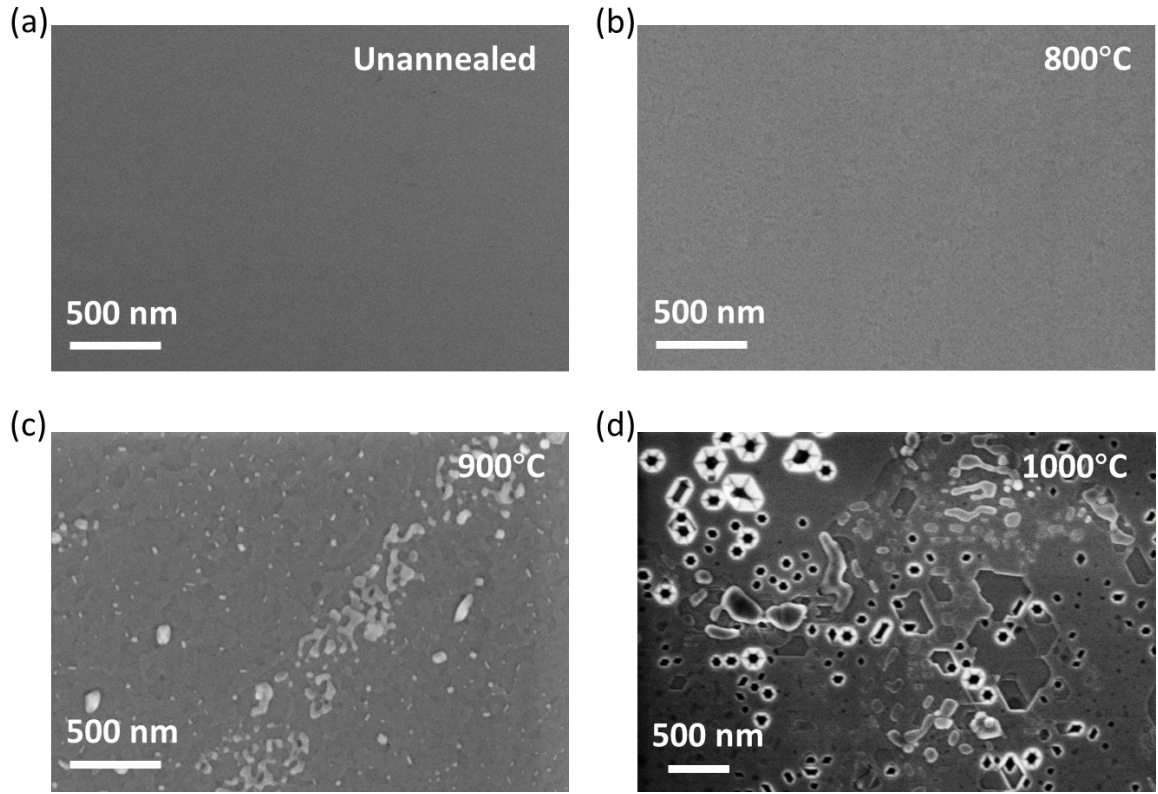

**Figure S7.** SEM images of ZnO areas irradiated with the ion fluence of $6\times10^{15}$ cm$^{-2}$ followed by **(a)** no annealing treatment, and **(b-d)** annealing in air for 40 mins at 800 °C, 900 °C, and 1000 °C, respectively.

**Section 5. Spectroscopic ellipsometry data acquisition and analysis of Ga:ZnO**

We performed infrared ellipsometry measurements (implemented using IR-VASE Mark II ellipsometer, J. A. Woollam) on Ga:ZnO samples for incident angles of 50°, 60°, and 70°. To extract refractive index of the material, a multi-layer model must be built to solve for the inverse problem: find the right n and κ and thickness of each layer to fit the measured Ψ and Δ. Note that in our modeling, the n and κ are correlated by the Kramers-Kronig relation and usually described by a series of optical oscillator functions such as Lorentz, Sellmeier, Drude, Gaussian, Cauchy, etc.

First, we characterized optical properties of the pristine ZnO substrate using a series of Gaussian oscillators (Table S1 and Fig. S7) expressed as [S9]:

$$\varepsilon_{Gaus} = \varepsilon_1 + i\varepsilon_2$$

$$\varepsilon_2 = A_n e^{\left(\frac{E-E_n}{\sigma}\right)^2} - A_n e^{-\left(\frac{E+E_n}{\sigma}\right)^2}$$

$$\varepsilon_1 = \frac{2}{\pi} P \int_0^\infty \frac{\xi \varepsilon_2(\xi)}{\xi^2 - E^2} d\xi$$

Where, $\sigma = \frac{Br_n}{2\sqrt{\ln(2)}}$, $Br_n$ = FWHM, P is the Cauchy Principal Value.

**Table S1.** Fitting parameters of Gaussian oscillators used for pristine ZnO substrate.

| Oscillator | Type | Amplitude $A_n$ | Center Energy $E_n$ (eV) | Broadening $Br_n$ (eV) |
|---|---|---|---|---|
| 1 | Gaussian | 85.165 | 0.0506 | 0.00210 |
| 2 | Gaussian | 16.737 | 0.05019 | 0.00610 |
| 3 | Gaussian | 0.32918 | 0.08231 | 0.03047 |
| 4 | Gaussian | 0.05138 | 0.17235 | 0.03051 |
| 5 | Gaussian | 0.21273 | 0.05917 | 0.01021 |
| 6 | Gaussian | 0.11585 | 0.19682 | 0.08180 |
| 7 | Gaussian | 0.03799 | 0.11703 | 0.01615 |

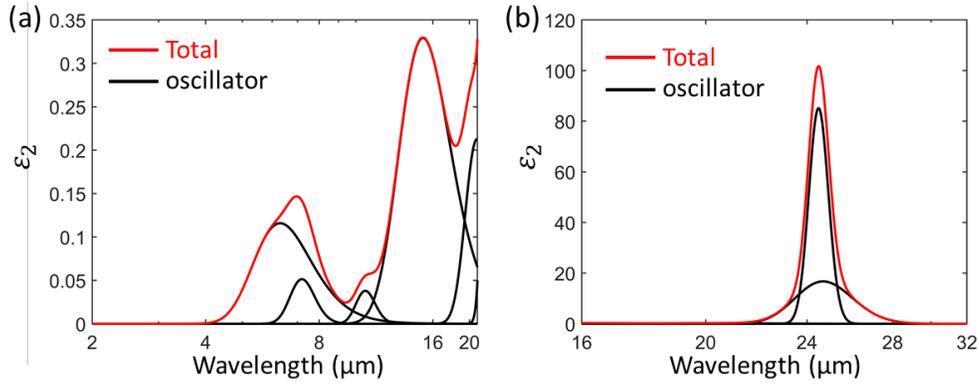

**Figure S8.** The fitting $\varepsilon_2$ values of the pristine ZnO substrate. The summed $\varepsilon_2$ function (red) and individual Gaussian oscillators (black) are plotted in **(b)** from 2 to 20 μm, and **(c)** from 16 to 32 μm.

As shown in Fig. S8 (a, b) the measured (discrete points) and fitted (continuous curves) Ψ and Δ for the pristine ZnO are in good agreement. Our modeling was also verified by comparing the calculated reflection based on our ellipsometry model to our FTIR reflectance measurements, as shown in Fig. S8(c). Fig. S8(d) plots the extracted complex refractive index of the pristine ZnO substrate.

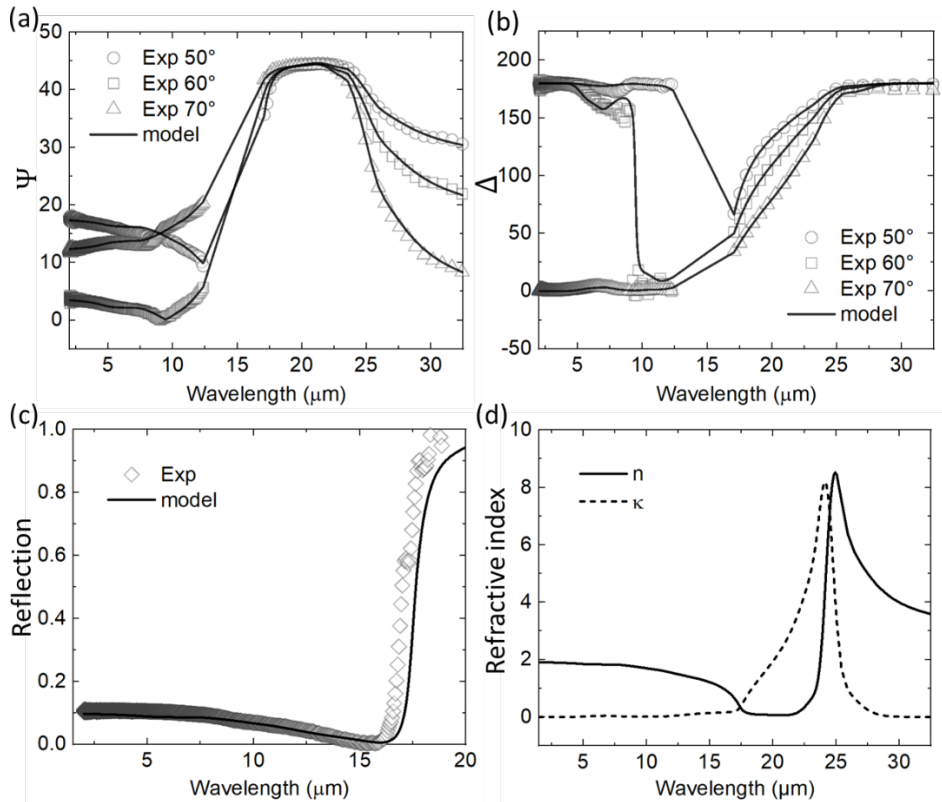

**Figure S9. (a, b)** Comparison between fitting (solid curves) and measured ellipsometry data of Ψ and Δ for the pristine ZnO substrate. **(c)** Comparison between calculated reflectance based on our ellipsometry fitting (solid curve) and FTIR measurements (diamond dots) for the pristine ZnO substrate. **(d)** Extracted complex refractive index of the pristine ZnO substrate.

Then, we followed the same fitting procedures to extract the optical properties of Ga:ZnO samples. For the thermal annealed Ga:ZnO samples, according to the SIMS depth profile shown in Fig. 2(d) in the main text, we found that there is a diffusion plateau underneath the top surface. Therefore, we built two layers to model the top surface and the diffusion layer separately, as discussed in the main text. Each layer includes an additional Drude function to account for the induced carrier concentration due to the FIB-assisted doping (note that the seven Gaussian oscillators are fixed, and only the Drude term was fitted.):

$$\varepsilon_{Drude} = \varepsilon_1 + i\varepsilon_2 = \frac{-\hbar^2}{\varepsilon_0 \rho_n (\tau_n \cdot E^2 + i\hbar E)}$$

$$\rho_n = \frac{m^*}{Nq^2\tau} = \frac{1}{q\mu N}$$

Where, $m^*$ is the effective mass, $\rho$ is the resistivity, and $\tau$ is the scattering time.

For the 3.1-at.% Ga-doped ZnO sample, we have six fitting parameters, including two sets of thickness, carrier concentration, and mobility: one set for the diffusion layer, and one set for the high carrier concentration layer. Fitting results and comparison between the measured and fitted $\Psi$ and $\Delta$ have been shown in Fig. 2 in the main text.

For the unannealed Ga:ZnO sample, we assumed a thin top layer (on the order of 10 nm, but the thickness was a fitting parameter) described by fixed Gaussian oscillators that describe intrinsic ZnO and an additional Drude term, with a semi-infinite undoped ZnO substrate. Note that this is similar to our fitting procedure with the annealed samples, but with no diffusion layer. The fitting results are shown below. Our fitting yielded a 20-nm layer with carrier concentration of ~$4.7 \times 10^{20}$ cm$^{-3}$ and a mobility of ~11.5 cm$^2$/V·s. Therefore, we observe a substantial carrier activation happening within this thin layer even in the absence of annealing, and the layer's optical properties can be described by a Drude model.

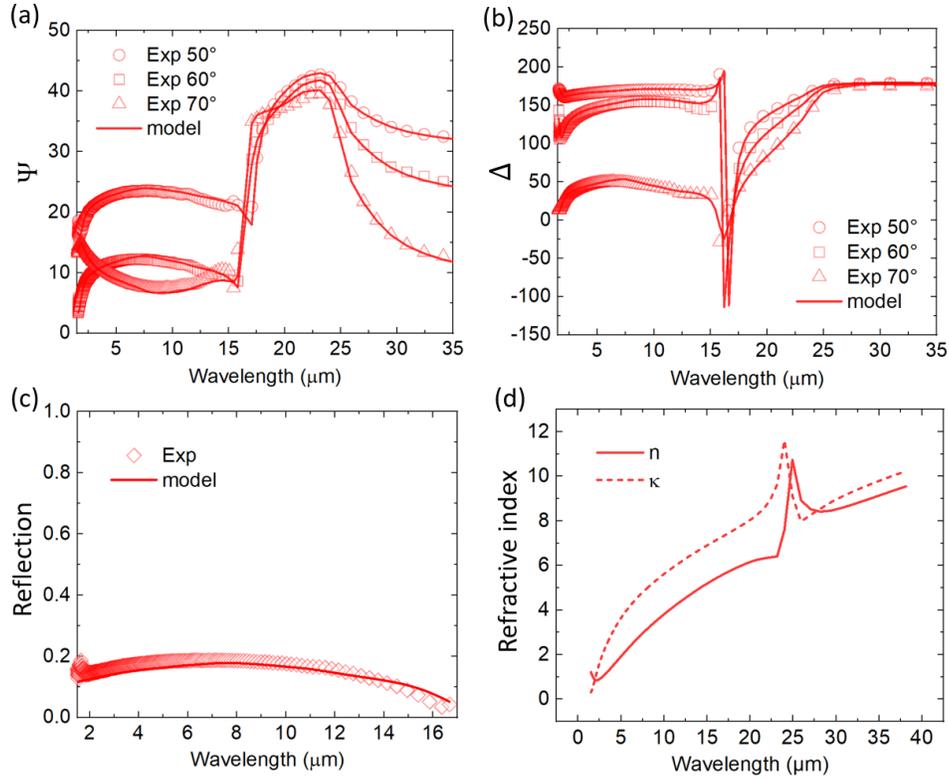

**Figure S10. (a, b)** Comparison between fitting (solid curves) and measured ellipsometry data of Ψ and Δ for the unannealed Ga:ZnO irradiated by a 30 keV ion accelerator with an ion fluence of 3.6 × 15 cm$^{-2}$ (corresponding to Ga peak concentrations of 3.1 at.%). **(c)** Comparison between calculated reflectance based on our ellipsometry fitting (solid curve) and FTIR measurements (diamond symbols) for the unannealed Ga:ZnO sample. **(d)** Extracted complex refractive index of the Ga:ZnO layer in the unannealed sample.

In the fitting process, a uniqueness test was performed to check the robustness of the fitting for different parameters. The ellipsometry fitting software WVASE™ uses the mean-squared error (MSE) as the figure of merit, which represents the quality of the match between the data calculated from the model and the experimental data. One way to examine whether the fitting of a parameter is accurate and unique is to take the best-fit model, slightly change the parameter of interest and fix its value (not fitting), and then refit all the other parameters around its value [S9].

Here the parameter of interest is the thickness of the top Ga:ZnO layer. Figure S11 shows the plot of the MSE values versus fixed thickness from 5 nm to 30 nm for unannealed sample, and samples annealed at 800 °C and 900 °C. We can observe that the thickness of 20 nm yields the minimum MSE for the unannealed Ga:ZnO, indicating a good agreement with the TRIM results. As shown in Figure S11(b,c), a thickness of 8 nm yields the minimum MSE for the samples annealed at 800 °C and 900 °C samples. This is a bit smaller than the thickness predicted by TRIM. We hypothesize that the top layer may become thinner due to diffusion of some of the dopants deeper into the sample, and we also suspect that the annealing treatment might push some of the dopants to the surface. However, it is also possible that the fitted thickness is imperfect because in reality there is a gradient of dopants through the material, rather than two homogeneous layers as assumed in the fit.

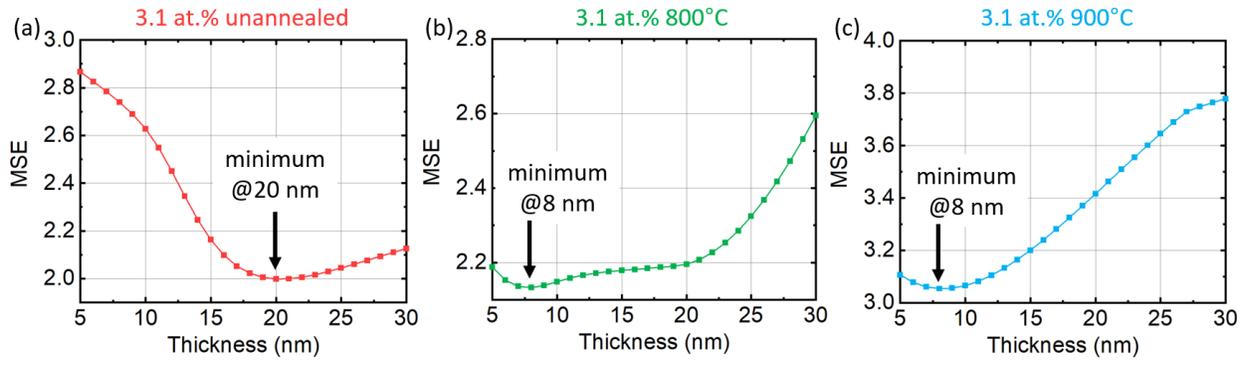

**Figure S11. (a-c)** Ellipsometry fitting uniqueness test for the thickness of the top layer of the unannealed Ga:ZnO, and for Ga:ZnO annealed at 800 °C and 900 °C, respectively.

**Section 6. Implantation parameters of FIB-VO$_2$ and SEM images**

We took SEM images of the FIB-VO$_2$ sample. As shown in Fig. S12, FIB-irradiated regions have a contrast compared to pristine regions, but there is no obvious change of the morphology caused by the FIB irradiation in comparison with the pristine VO$_2$ region, as seen in the high-resolution SEM Figure S12(e).

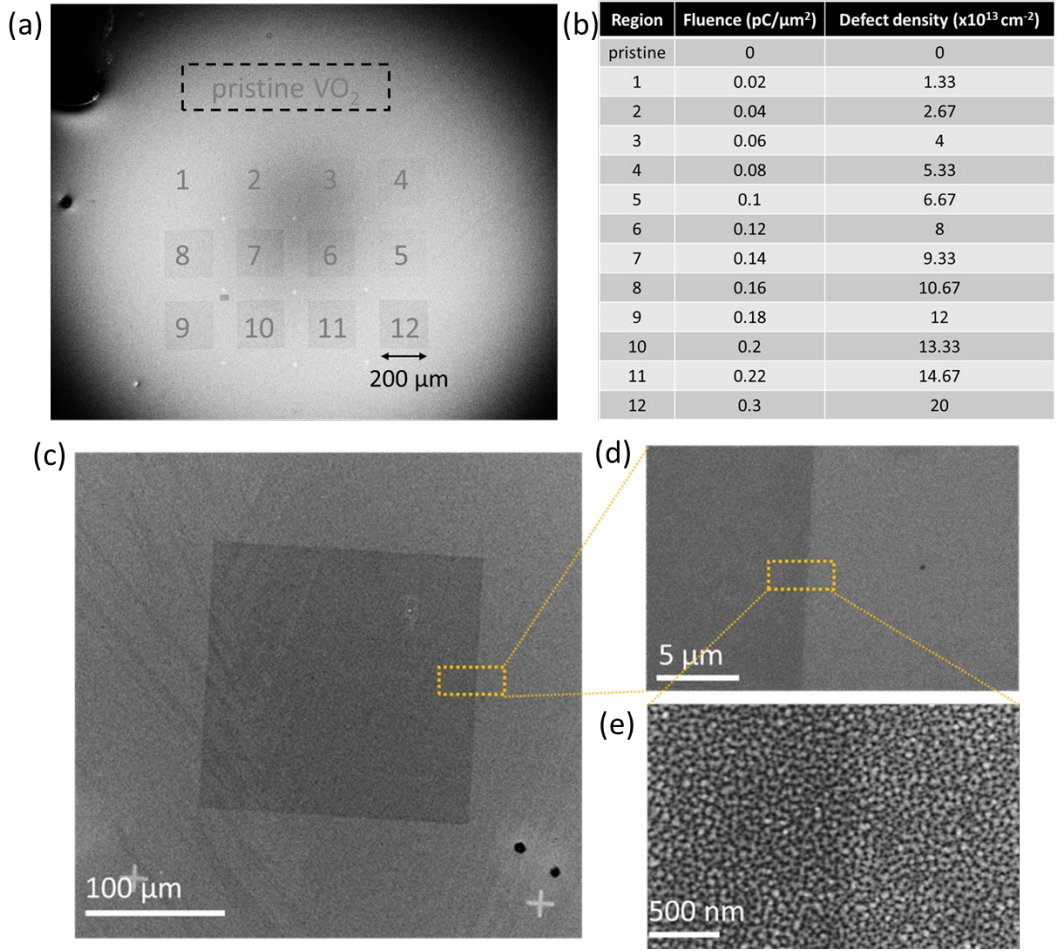

**Figure S12. (a,b)** SEM image of the top surface of the FIB-VO$_2$ sample with the corresponding ion fluences given in the table. **(c-e)** SEM images of the FIB-irradiated VO$_2$ region #12 with an ion fluence of 2 ×14 cm$^{-2}$.

We performed Raman mapping to show that FIB is able to locally modulate the IMT of VO$_2$ within sub-micrometer areas. We used an excitation laser of 532 nm with a step size of 1 µm of a 5-by-40-µm rectangular region that included both pristine and irradiated regions. As shown in Fig. S13, the irradiated VO$_2$ region was transformed to the metallic phase at 50 °C, featuring no Raman mode at 610 cm$^{-1}$, while the pristine region was still in its insulating phase, with a strong Raman mode at 610 cm$^{-1}$.

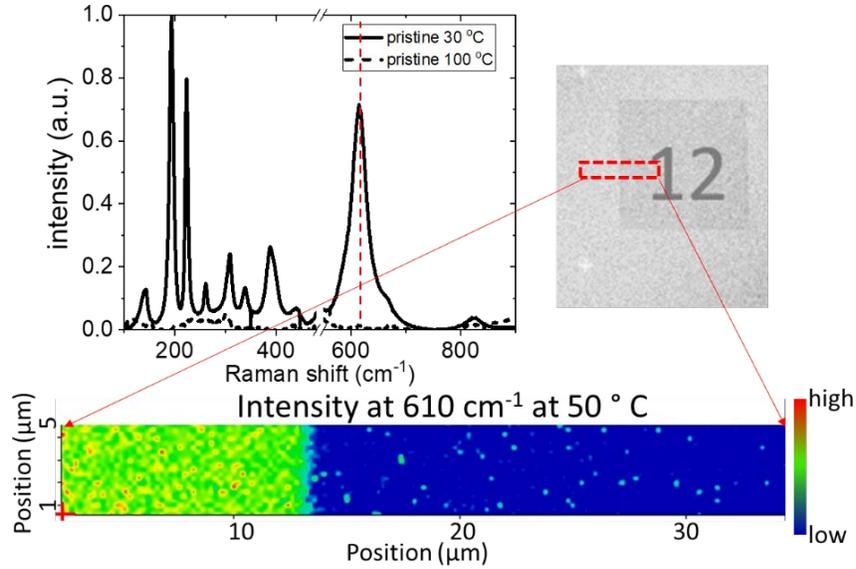

**Figure S13.** Raman mapping across the boundary between a pristine region and a region irradiated with fluence of 2 × 14 cm$^{-2}$.

We also took SEM images of cross section of the FIB-VO$_2$ sample. The cross section was created by a FIB-assisted milling process (implemented using FIB-SEM, Zeiss Auriga), as shown in Fig S14. In order to obtain clear imaging of the VO$_2$ boundaries, we pre-deposited ~300-nm thick, ~1-by-1-mm patch of copper (Cu) on top of the VO$_2$ and milled through the layers within this Cu patch area, resulting in a clean cross section of the VO$_2$ layer for thickness measurement. The thickness of VO$_2$ was measured to be (52 ± 5) nm, after we compensated the angle of 54° between the FIB and SEM beam.

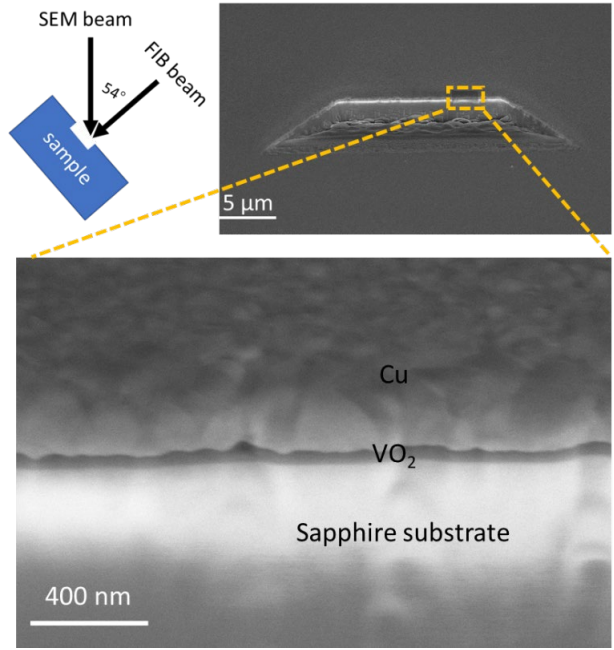

**Figure S14.** Schematic of FIB-milling-assisted cross-section imaging of the FIB-VO$_2$ sample. The measured thickness of VO$_2$ is (52±5) nm.

**Section 7. Full dataset of optical refractive indices of FIB-VO$_2$ for different ion fluences**

In Fig. 4(c) of the main text, we only plotted the characterized refractive indices of the FIB-VO$_2$ for a single wavelength of 9 μm to clearly show the evolution of refractive-index values versus temperature and ion fluence. Fig. S15 includes refractive indices of VO$_2$ FIB-irradiated by different fluences, for wavelengths of 6 – 14 μm.

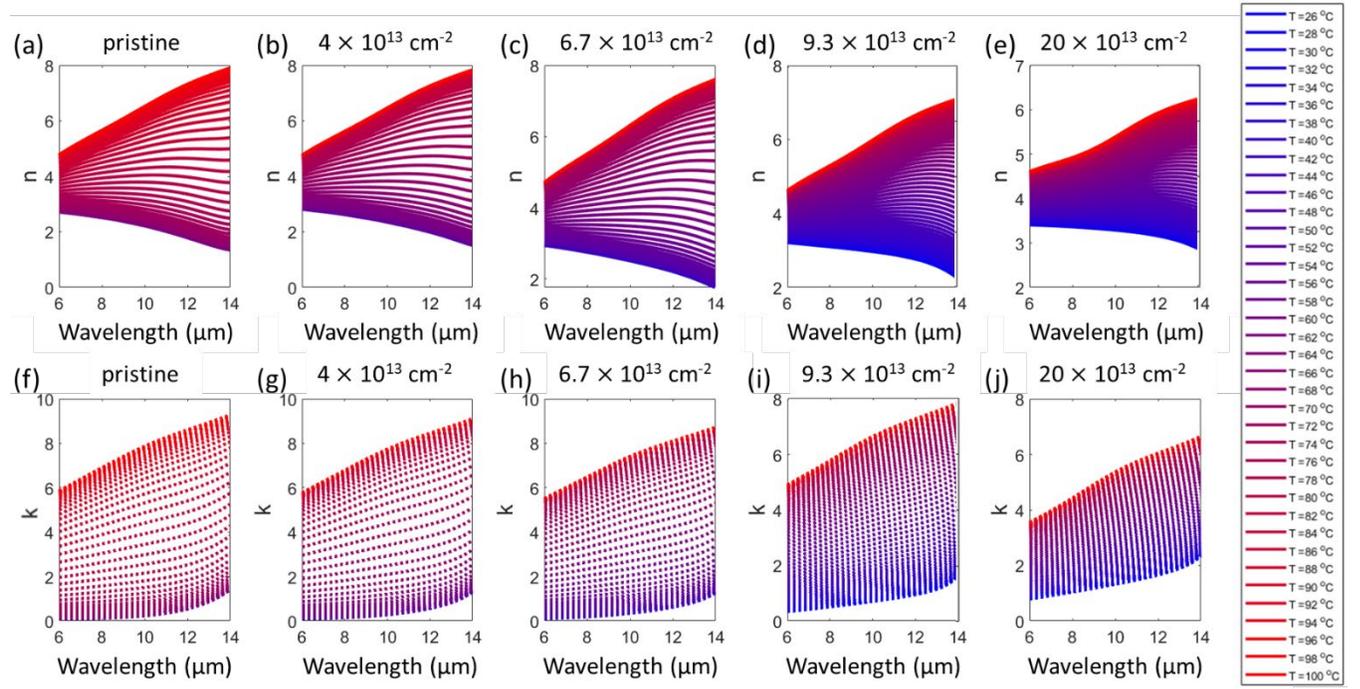

**Figure S15.** Full dataset of **(a-e)** real and **(f-j)** imaginary parts of temperature-dependent refractive indices of VO$_2$ irradiated with different ion fluences as indicated on top of each figure, for wavelengths from 6 to 14 μm.

**Section 8. Stability of the hysteresis curves after multiple thermal cycles**

To check the stability of the hysteresis curves after multiple thermal cycles, we ran temperature-dependent FTIR reflectance measurements on region #6 (irradiated with the ion fluence of $8\times13$ cm$^{-2}$) of our FIB-VO$_2$ sample for 3 cycles of heating and cooling, as shown in the figure below. Our results show that the hysteresis and IMT temperature is stable with multiple thermal cycles between room temperature and 90 °C.

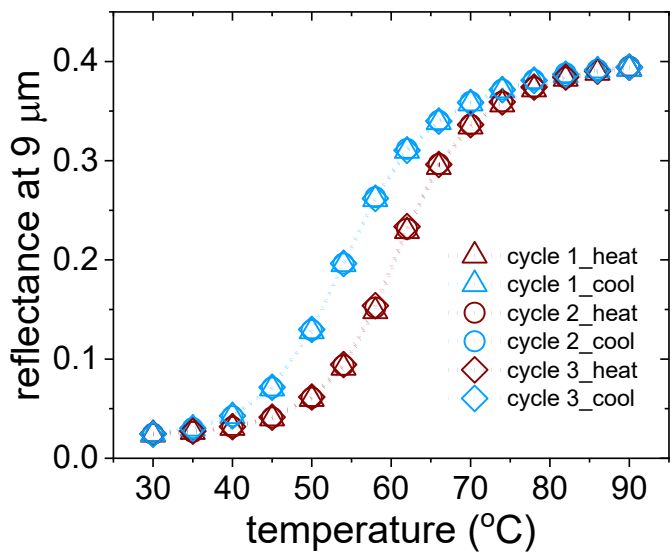

**Figure S16.** FTIR reflectance measurements on FIB-VO$_2$ for 3 cycles of heating and cooling between room temperature and 90 °C.